\documentclass[letterpaper ,10pt, conference]{IEEEtran}      

\IEEEoverridecommandlockouts                              

\date{}

\usepackage{graphics} 
\usepackage{epsfig} 
\usepackage{amsmath} 
\usepackage{amssymb}  
\usepackage{amssymb,amsmath}
\usepackage{graphicx,color}

\newcommand{\prob}{{\mathsf P}}

\newcommand{\tends}{\rightarrow}

\newcommand{\qed}{\hspace*{\fill}~{\rule{2mm}{2mm}}\par\endtrivlist\unskip}

\newcommand{\expec}{{\mathbb E}}

\newcommand{\vect}[1]{{\pmb #1}}

\newcommand{\snr}{{\mathsf{SNR}}}

\newtheorem{lemma}{Lemma}
\newtheorem{theorem}{Theorem}

\newtheorem{proposition}{Proposition}

\begin{document}

\title{Capacity of the Discrete-Time AWGN Channel Under Output Quantization}

\author{Jaspreet Singh, Onkar Dabeer and Upamanyu Madhow$^*$
\thanks{}
\thanks{$^*$J. Singh and U. Madhow are with the ECE Department, UC Santa Barbara, CA
93106, USA. Their research was supported by the National Science
Foundation under grant CCF-0729222 and by the Office for Naval
Research under grant N00014-06-1-0066. O. Dabeer is with the
Tata Institute of Fundamental Research, Mumbai 400005, India.
His work was supported in part by a grant from the Dept. of
Science and Technology, Govt.\ of India, and in part by the Homi
Bhabha Fellowship.
{\tt\small  \{jsingh, madhow\}@ece.ucsb.edu, onkar@tcs.tifr.res.in}}%
}

\maketitle

\begin{abstract}
We investigate the limits of communication over the
discrete-time Additive White Gaussian Noise (AWGN) channel, when
the channel output is quantized using a small number of bits. We
first provide a proof of our recent conjecture on the optimality
of a discrete input distribution in this scenario. Specifically,
we show that for any given output quantizer choice with $K$
quantization bins (i.e., a precision of $\log_2{K}$ bits), the
input distribution, under an {\it average power} constraint,
need not have any more than $K+1$ mass points to achieve the
channel capacity. The cutting-plane algorithm is employed to
compute this capacity and to generate optimum input
distributions. Numerical optimization over the choice of the
quantizer is then performed (for $2$-bit and $3$-bit symmetric
quantization), and the results we obtain show that the loss due
to low-precision output quantization, which is small at low
signal-to-noise ratio ($\snr$) as expected, can be quite
acceptable even for moderate to high $\snr$ values. For example,
at $\snr$s up to $20$ dB, $2$-$3$ bit quantization achieves
$80$-$90 \%$ of the capacity achievable using infinite-precision
quantization.
\end{abstract}

\section{Introduction} \label{sec:intro}
Analog-to-digital conversion (ADC) is an integral part of modern
communication receiver architectures based on digital signal
processing (DSP). Typically, ADCs with $6$-$12$ bits of
precision are employed at the receiver to convert the received
analog baseband signal into digital form for further processing.
However, as the communication systems scale up in speed and
bandwidth (for e.g., systems operating in the ultrawide band or
the mm-wave band), the cost and power consumption of such high
precision ADC becomes prohibitive \cite{Walden}. A DSP-centric
architecture nonetheless remains attractive, due to the
continuing exponential advances in digital electronics (Moore's
law). It is of interest, therefore, to understand whether
DSP-centric design is compatible with the use of low-precision
ADC.

\begin{figure}[]
  \centering
    \begin{picture}(131,76.9)
 \put(-28,6.7){\scalebox{.33}{\includegraphics{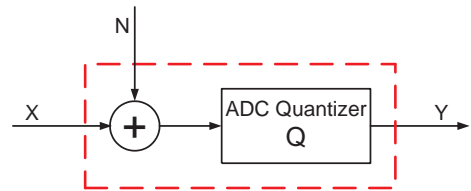}}}
 \end{picture}
  \caption{The \emph{AWGN-Quantized Ouput} Channel  : $Y = \mathsf{Q}(X+N)$ . }\label{fig:Channel}
\end{figure}

In this paper, we continue our investigation of the
Shannon-theoretic communication limits imposed by the use of
low-precision ADC for ideal Nyquist sampled linear modulation in
AWGN. The discrete-time memoryless AWGN-{\it Quantized Output}
(AWGN-QO) channel model thus induced is shown in
Fig.~\ref{fig:Channel}. In our prior work for this channel
model, we have shown that for the extreme scenario of 1-bit
symmetric quantization, binary antipodal signaling achieves the
channel capacity for any signal-to-noise ratio ($\snr$)
\cite{Spawc06}. For multi-bit quantization \cite{ICC07}, we
provided a duality-based approach to bound the capacity from
above, and employed the cutting-plane algorithm to generate
input distributions that nearly achieved these upper bounds.
Based on our results, we conjectured that a discrete input with
cardinality not exceeding the number of quantization bins
achieves the capacity of the {\it average power} constrained
AWGN-QO channel. In this work, we prove that a discrete input is
indeed optimal, although our result only guarantees its
cardinality to be {\it at most} $K+1$, where $K$ is the number
of quantization bins. Our proof is inspired by Witsenhausen's
result in \cite{Witsenhausen}, where Dubins' theorem
\cite{Dubins} was used to show that the capacity of a
discrete-time memoryless channel with output cardinality $K$,
under only a {\it peak power} constraint is achievable by a
discrete input with at most $K$ points. The key to our proof is
to show that, under output quantization, an average power
constraint automatically induces a peak power constraint, after
which we use Dubins' theorem as done by Witsenhausen. Although
not applicable to our setting, it is worth noting that for a
Discrete Memoryless Channel, Gallager first showed that the
number of inputs with nonzero probability mass need not exceed
the number of outputs \cite[p. 96, Corollary 3]{Gallager}. 

While the preceding results optimize the input distribution for
a fixed quantizer, comparison with an unquantized system
requires optimization over the choice of the quantizer as well.
We do this numerically for $2$-bit and $3$-bit symmetric
quantization, and use our numerical results to make the
following encouraging observations: (a) Low-precision ADC incurs
a relatively small loss in spectral efficiency compared to
unquantized observations. While this is expected for low
$\snr$s, we find that even at moderately high $\snr$s of up to
$20$ dB, $2$-$3$ bit ADC still achieves 80-90$\%$ of the
spectral efficiency attained using unquantized observations.
These results indicate the feasibility of system design using
low-precision ADC for high bandwidth systems. (b) Standard
uniform Pulse Amplitude Modulated (PAM) input with quantizer
thresholds set to implement maximum likelihood (ML) hard
decisions achieves nearly the same performance as that attained
by an optimal input and quantizer pair. This is useful from a
system designer's point of view, since the ML quantizer
thresholds have a simple analytical dependence on SNR, which is
an easily measurable quantity.

The rest of the paper is organized as follows. The quantized
output AWGN channel model is given in the next section. In
Section \ref{sec:Properties}, we show that a discrete input
achieves the capacity of this channel. Quantizer optimization
results are presented in Section \ref{sec:Quantizer}, followed
by the conclusions in Section \ref{sec:Conclusions}.

\section{Channel Model}\label{sec:Channel Model}
We consider linear modulation over a real AWGN channel, and
assume that the Nyquist criterion for no intersymbol
interference is satisfied \cite[pp. 50]{Madhow_book}.  Symbol
rate sampling of the receiver's matched filter output using a
finite-precision ADC therefore results in the following
discrete-time memoryless AWGN-\emph{Quantized Output} (AWGN-QO)
channel (Fig. 1)
\begin{equation}\label{eq:Channel1}
Y= \mathsf{Q}\left(X+ N\right) \ .
\end{equation}
Here $X \in \mathbb{R}$ is the channel input with distribution
$F(x)$ and $N$ is ${\cal N}(0,\sigma^2)$. The quantizer
$\mathsf{Q}$ maps the real valued input $X+N$ to one of the $K$
bins, producing a discrete channel output $Y \in \{y_1, \cdots,
y_K \}$. We only consider quantizers for which each bin is an
interval of the real line. The quantizer $\mathsf{Q}$ with $K$
bins can therefore be characterized by the set of its $(K-1)$
thresholds $\vect{q}=[q_1, q_2, \cdots, q_{K-1}] \in
\mathbb{R}^{K-1}$, such that $-\infty := q_0 < q_1 < q_2 <
\cdots < q_{K-1} < q_K := \infty$. The resulting transition
probability functions are given by
\begin{equation} \label{eq:Transitions}
W_i(x)=\prob(Y=y_i|X=x)=Q\left(\frac{q_{i-1}-x}{\sigma}\right)-Q\left(\frac{q_{i}-x}{\sigma}\right)
\ ,
\end{equation}
where $Q(x)$ denotes the complementary Gaussian distribution
function $\frac{1}{\sqrt{2 \pi}}\int_{x}^{\infty} \exp(-t^2/2)
dt$.

The input-output mutual information $I(X;Y)$, expressed
explicitly as a function of $F$ is
\begin{equation}\label{eq:MutualInfo}
I(F)=\int_{-\infty}^{\infty}
{\sum_{i=1}^{K}{W_i(x)\log{\frac{W_i(x)}{R(y_i;F)}}}}dF(x) \ ,
\end{equation}
where $\{R(y_i;F) \ , 1\leq i\leq K \}$ is the Probability Mass
Function (PMF) of the output when the input is $F$.
Under an average power constraint $P$ (i.e., $\expec[X^2] \leq P$),
we wish to compute the capacity of the channel \eqref{eq:Channel1},
which is given by
\begin{equation}\label{eq:Capacity1}
C=\sup_{F \in \cal{F}}{I(F)},
\end{equation}
where $\cal{F}$ is the set of all average power constrained
distributions on $\mathbb{R}$.

\section{Discrete Input Achieves Capacity}\label{sec:Properties} We
first use the Karush-Kuhn-Tucker (KKT) optimality condition to
show that an average power constraint for the AWGN-QO channel
automatically induces a constraint on the peak power, in the
sense that an optimal input distribution must have a bounded
support set. This fact is then exploited to show the optimality
of a discrete input.
\subsection{An Implicit peak power Constraint}\label{sec:Peak}
The following KKT condition can be derived for the AWGN-QO
channel, using convex optimization principles in a manner
similar to that in \cite{Smith_D, Faycal}.
The input distribution $F$ is optimal if and only if there
exists a $\gamma \geq 0$ such that
\begin{equation}\label{eq:KKT2}
\sum_{i=1}^{K}{W_i(x) \log{\frac{W_i(x)}{R(y_i;F)}}} +
\gamma(P-x^2) \leq I(F) \ ,
\end{equation}
for all $x$, with equality if $x$ is in the support of $F$.

The first term on the left hand side of the KKT condition
\eqref{eq:KKT2} 
is the divergence (or the relative entropy) between the
transition and the output PMFs. For convenience, let us denote
it by $d(x;F)$. The following result concerning the behavior of
$d(x;F)$ has been proved in \cite{JP_Draft}.
\begin{lemma}\label{Lemma:behavior}
For the AWGN-QO channel \eqref{eq:Channel1} with input
distribution $F$, the divergence
function $d(x;F)$ satisfies the following properties \\
(a)$\displaystyle\lim_{x \to \infty} d(x;F)=-\log{R(y_K;F)}$.\\
(b) There exists a finite constant $A_0$ such that $\forall \
x>{A_0}, \newline  d(x;F) < -\log{R(y_K;F)}$.
\end{lemma}
\proof See \cite{JP_Draft}.

%

We now use Lemma \ref{Lemma:behavior} to prove the main result
of this subsection.
\begin{proposition} A capacity-achieving input distribution
for the average power constrained AWGN-QO channel
\eqref{eq:Channel1} must have bounded support.
\end{proposition}
{\it Proof:} Assume that the input distribution $F^*$
achieves\footnote{That the capacity is achievable can be shown
using standard results from optimization theory. For lack of
space here, we refer the reader to \cite{JP_Draft} for details.}
the capacity in \eqref{eq:Capacity1} (i.e., $I(F^*)=C$), with
$\gamma^* \geq 0$ being a corresponding optimal Lagrange
parameter in the KKT condition. In other words, with
$\gamma=\gamma^*$, and, $F=F^*$, (\ref{eq:KKT2}) must be
satisfied with an equality at every point in the support of
$F^*$. We exploit this necessary condition next to show that the
support of $F^*$ is upper bounded. Specifically, we prove that
there exists a finite constant ${A_2}^*$ such that it is not
possible to attain equality in (\ref{eq:KKT2}) for any
$x>{A_2}^*$.

Using Lemma $1$, we first let \newline $\displaystyle\lim_{x \to
\infty} d(x;F^*)= -\log(R(y_K;F^*)) = L$, and also assume that
there exists a finite constant $A_0$ such that $\forall \
x>{A_0}$, $d(x;F^*) < L$. We consider two possible cases.
\begin{itemize}
\item Case 1: $\gamma^*>0$.\\
If $C > L + \gamma^\ast P$, then pick ${A_2}^* = A_0$. \\
Else pick ${A_2}^* \geq \max\{A_0, \sqrt{(L+\gamma^\ast
P-C)/\gamma^\ast}\}$.\\
In either situation, for $x > {A_2}^*$, we get $d(x;F^*) < L$,
and, $\gamma^\ast x^2
> L  + \gamma^* P - C $.\\
This gives
\[
d(x;F^*) + \gamma^*(P-x^2) < L + \gamma^*P - (L + \gamma^* P -
C) = C .
\]
\item Case 2: $\gamma^*=0$. \\
Putting $\gamma^*=0$ in the KKT condition \eqref{eq:KKT2}, we
get
\begin{equation*}
d(x;F^*) = \sum_{i=1}^{K}{W_i(x)
\log{\frac{W_i(x)}{R(y_i;F^*)}}} \leq C \ , \ \  \forall x .
\end{equation*}
Thus,
\begin{equation*}
L = \displaystyle\lim_{x \to \infty} d(x;F^*) \leq C .
\end{equation*}
Picking ${A_2}^* = A_0$, we therefore have that for $x >
{A_2}^*$
\begin{equation*}
\begin{split}
& d(x;F^*) + \gamma^*(P-x^2) = d(x;F^*) < L .\\
\implies & d(x;F^*) + \gamma^*(P-x^2) < C .
\end{split}
\end{equation*}
\end{itemize}
Combining the two cases, we have shown that the support of the
distribution $F^*$ has a finite upper bound ${A_2}^*$. Using
similar arguments, it can easily be shown that the support of
$F^*$ has a finite lower bound ${A_1}^*$ as well, which implies
that $F^*$ has a bounded support.\qed

\subsection{Achievability of Capacity by a Discrete
Input}\label{sec:Discrete} To show the optimality of a discrete
input for our problem, we use the following theorem which we
have proved in \cite{JP_Draft}. The theorem holds for channels
with a finite output alphabet, under the condition that the
input is constrained in both peak power and average power.
\begin{theorem}
Consider a stationary discrete-time memoryless channel with a
continuous input $X$ taking values in the bounded interval
$[A_1, A_2]$, and a discrete output $Y \in \{y_1, y_2, \cdots,
y_K\}$. Let the transition probability function
$W_i(x)=\prob(Y=y_i|X=x)$ be continuous in $x$, for each $i$ in
$\{1,..,K\}$. The capacity of this channel, under an average
power constraint on the input, is achievable by a discrete input
with at most $K+1$ points.
\end{theorem}
{\it Proof:} See \cite{JP_Draft}. \qed
Our proof in \cite{JP_Draft} uses Dubins' theorem \cite{Dubins}, and
is an extension of Witsenhausen's result in \cite{Witsenhausen},
wherein he showed that a distribution with only $K$ points would be
sufficient to achieve the capacity if the average power of the input
was not constrained.

The implicit peak power constraint derived in Section
\ref{sec:Peak} allows us to use Theorem $1$ to get the following
result.
\begin{proposition}
The capacity of the average power constrained AWGN-QO channel
\eqref{eq:Channel1} is achievable by a discrete input distribution
with at most $K+1$ points of support.
\end{proposition}
{\it Proof:} Using notation from the last subsection, let $F^*$ be
an optimal distribution for \eqref{eq:Capacity1}, with the support
of $F^*$ being contained in the bounded interval $[{A_1}^*,
{A_2}^*]$. Define $\mathcal{F}_1$ to be the set of all average power
constrained distributions whose support is contained in $[{A_1}^*,
{A_2}^*]$.
Note that $F^* \in {\mathcal{F}_1} \subset \cal{F}$, where $\cal{F}$
is the set of all average power constrained distributions on
$\mathbb{R}$. Consider the maximization of the mutual information
$I(X;Y)$ over the set ${\mathcal{F}_1}$
\begin{equation}\label{eq:Capacity2}
C_1=\max_{F \in \mathcal{F}_1} I(F).
\end{equation}
Since the transition probability functions in
\eqref{eq:Transitions} are continuous in $x$, Theorem $1$
implies that a discrete distribution with at most $K+1$ mass
points achieves the maximum $C_1$ in \eqref{eq:Capacity2}.
Denote such a distribution by $F_1$. However, since $F^*$
achieves the maximum $C$ in \eqref{eq:Capacity1} and $F^{\ast}
\in {\mathcal{F}_1}$, it must also achieve the maximum in
\eqref{eq:Capacity2}. This implies that $C_1=C$, and that $F_1$
is optimal for \eqref{eq:Capacity1}, thus completing the
proof.\qed
\subsection{Capacity Computation}\label{sec:Fixed}
We have already addressed the issue of computing the capacity
\eqref{eq:Capacity1} in our prior work. Specifically, in
\cite{Spawc06}, we have shown analytically that for the extreme
scenario of 1-bit symmetric quantization, binary antipodal
signaling achieves the capacity (at any $\snr$). Multi-bit
quantization has been considered in \cite{ICC07, JP_Draft},
where we show that the cutting-plane algorithm \cite{Huang-Meyn}
can be employed for computing the
capacity and obtaining optimal input distributions. 

\section{Optimization Over Quantizer} \label{sec:Quantizer}
Until now, we have addressed the problem of capacity computation
given a fixed quantizer. In this section, we consider the issue
of quantizer optimization, while restricting attention to {\it
{symmetric quantizers only}}. Given the symmetric nature of the
AWGN noise and the power constraint, it seems intuitively
plausible that restriction to symmetric quantizers should not be
sub-optimal from the point of view of optimizing over the
quantizer choice in \eqref{eq:Channel1}, although a proof of
this conjecture has eluded us.

{\emph{A Simple Benchmark:}} While an optimal quantizer (with a
corresponding optimal input) provides the absolute communication
limits for our model, from a system designer's perspective, it
would also be useful to evaluate the performance degradation if
we use some standard input constellations and quantizer choices.
We take the following input and quantizer pair as our {\it
benchmark strategy} : for K-bin quantization, consider
equispaced uniform K-PAM (Pulse Amplitude Modulated) input
distribution, with quantizer thresholds as the mid-points of the
input mass point locations (i.e., ML hard decisions). With the
$K$-point uniform input, we have the entropy $H(X)=\log_2{K}$
bits for any $\snr$. Also, it is easy to see that as $\snr
\tends \infty$, $H(X|Y) \tends 0$ for the benchmark
input-quantizer pair. Therefore, our benchmark scheme is
near-optimal if we operate in the high $\snr$ regime. The main
issue to investigate ahead, therefore is: at low to moderate
$\snr$s, how much gain does an optimal quantizer choice provide
over the benchmark.

In all the results that follow, we take the noise variance
$\sigma^2=1$. However, the results are scale invariant in the
sense that if both $P$ and $\sigma^2$ are scaled by the same
factor $R$ (thus keeping the $\snr$ unchanged), then there is an
equivalent quantizer (obtained by scaling the thresholds by
$\sqrt{R}$) that gives an identical performance.

\section*{Numerical Results}

\subsection{$2$-bit Symmetric Quantization}
\begin{figure}[tb]
  \centering
    \begin{picture}(165,170.9)
 \put(-47,-10){\scalebox{.62}{\includegraphics{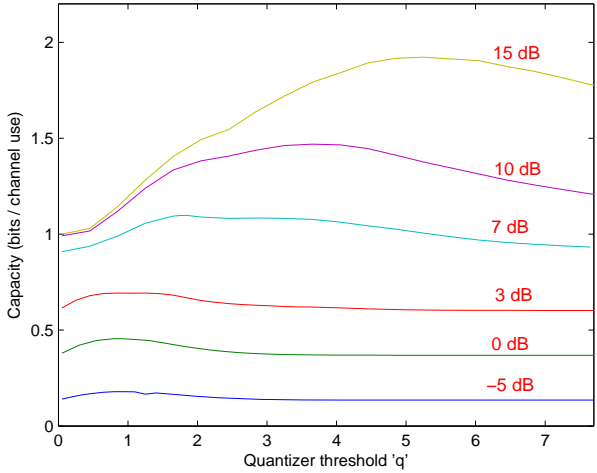}}}
 \end{picture}
  \caption{2-bit symmetric quantization : channel capacity versus the quantizer threshold $q$ (noise variance $\sigma^2 =
1$).}\label{fig:v_qp}
\end{figure}

A $2$-bit symmetric quantizer is characterized by a single
parameter $q$, with $\{-q, 0, q\}$ being the quantizer
thresholds. Hence we use a brute force search over $q$ to
optimize the quantizer. In Fig.~\ref{fig:v_qp}, we plot the
variation of the channel capacity (computed using the
cutting-plane algorithm) as a function of the parameter $q$ at
various $\snr$s. We observe that for any $\snr$, there is an
optimal choice of $q$ that maximizes the capacity. At high
$\snr$s, the optimal $q$ is seen to increase monotonically with
$\snr$, which is not surprising since the benchmark quantizer's
$q$ scales as $\sqrt{\snr}$ and is known to be near-optimal at
high $\snr$s.
\begin{table}
\centering
\begin{tabular}{|c|c|c|c|c|c|c|c|c|}
\hline
$\snr$(dB) & $-10$ & $-5$ & $0$ & $7$ & $15$ \\
\hline $1$-bit optimal & $0.0449$ & $0.1353$ & $0.3689$ & $0.9020$ & $0.9974$ \\
\hline $2$-bit optimal & $0.0613$ & $0.1792$ & $0.4552$ &  $1.0981$ & $1.9304$ \\
\hline $2$-bit benchmark & $0.0527$ & $0.1658$ & $0.4401$ &  $1.0639$ & $1.9211$ \\
\hline
\end{tabular}
\caption{Mutual information (in bits/channel use) at different
$\snr$s.}\label{Tab_2bit}
\end{table}

{\emph{Comparison with the benchmark:}} In Table \ref{Tab_2bit},
we compare the performance of the optimal solution obtained as
above with the benchmark scheme. The capacity with 1-bit
quantization is also shown for reference. While being
near-optimal at moderate to high $\snr$s, the benchmark scheme
is seen to perform fairly well at low $\snr$s also. For
instance, at $-10$ dB $\snr$, it achieves $86 \%$ of the
capacity achieved with an optimal $2$-bit quantizer and input
pair. From a practical standpoint, these results imply that the
benchmark scheme, which requires negligible computational effort
(due to its well-defined dependence on $\snr$), can be employed
even at small $\snr$s while incurring an acceptable loss of
performance.

\begin{figure}[tb]
  \centering
    \begin{picture}(92,128)
 \put(-105,-16){\scalebox{.41}{\includegraphics{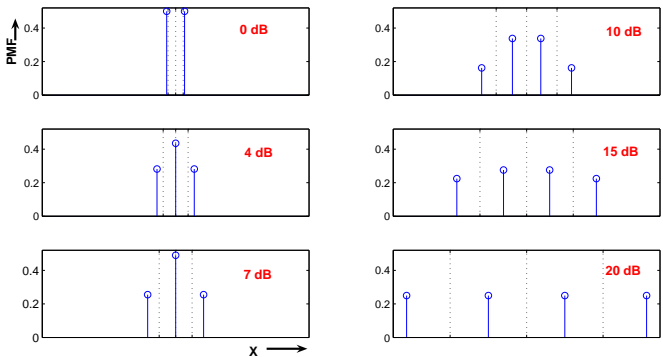}}}
 \end{picture}
  \caption{$2$-bit symmetric quantization : optimal input distribution and quantizer at various $\snr$s
  (the dashed vertical lines depict the locations of the quantizer
  thresholds).
}\label{fig:2-bit}
\end{figure}

{\emph{Optimal Input Distributions}}: The optimal input
distributions (given by the cutting-plane algorithm)
corresponding to the optimal quantizers obtained above are
depicted in Fig. \ref{fig:2-bit}, for different $\snr$ values.
The locations of the optimal quantizer thresholds are also shown
(by the dashed vertical lines). Binary signaling is found to be
optimal at low $\snr$s, and the number of mass points increases
(first to $3$ and then to $4$) with increasing $\snr$. Further
increase in $\snr$ eventually leads to the uniform $4$-PAM
input, thus approaching the capacity bound of $2$ bits. It is
worth noting that all the optimal inputs we obtained have $4$ or
less mass points, whereas Proposition $2$ is looser as it
guarantees the achievability of capacity using at most $5$
points.


\subsection{3-bit Symmetric Quantization}
For $3$-bit symmetric quantization, we need to optimize over a
space of $3$ parameters : $\{0 < q_1 < q_2 < q_3\}$, with the
quantizer thresholds being $\{\pm q_1, \pm q_2, \pm q_3\}$.
Instead of brute force search, we use an alternate optimization
procedure for joint optimization of the input and the quantizer
in this case. Due to lack of space, we refer the reader to
\cite{JP_Draft} for details, and proceed directly to the
numerical results. (Table \ref{Tab_3bit})
\begin{table}
\centering
\begin{tabular}{|c|c|c|c|c|c|c|c|c|}
\hline
$\snr$(dB) & $-10$ & $0$ & $5$ & $10$ & $20$ \\
\hline $3$-bit optimal & $0.0667$ & $0.4817$ &  $0.9753$ & $1.5844$ & $2.8367$ \\
\hline $3$-bit benchmark & $0.0557$ & $0.4707$ & $0.9547$ &  $1.5332$ & $2.8084$ \\
\hline
\end{tabular}
\caption{Mutual information (in bits/channel use) at different
$\snr$s.}\label{Tab_3bit}
\end{table}

{\emph{Comparison with the benchmark:}} As for $2$-bit quantization
considered earlier, we find that the benchmark scheme performs quite
well at low $\snr$s with $3$-bit quantization also. At $-10$ dB
$\snr$, for instance, the benchmark scheme achieves $83 \%$ of the
capacity achievable with an optimal quantizer choice. Table
\ref{Tab_3bit} gives the comparison for different $\snr$s.

{\emph{Optimal Input Distributions}}: Although not depicted
here, we again observe (as for the $2$-bit case) that the
optimal inputs obtained all have at most $K$ points ($K=8$ in
this case), while Proposition $2$ guarantees the achievability
of capacity by at most $K+1$ points. Of course, Proposition $2$
is applicable to any quantizer choice (and not just optimal
symmetric quantizers that we consider in this section), it still
leaves us with the question whether it can be tightened to
guarantee achievability of capacity with at most $K$ points.
\subsection{Comparison with Unquantized Observations}
\begin{figure}[tb]
  \centering
    \begin{picture}(144,160.9)
 \put(-53,-12){\scalebox{.6}{\includegraphics{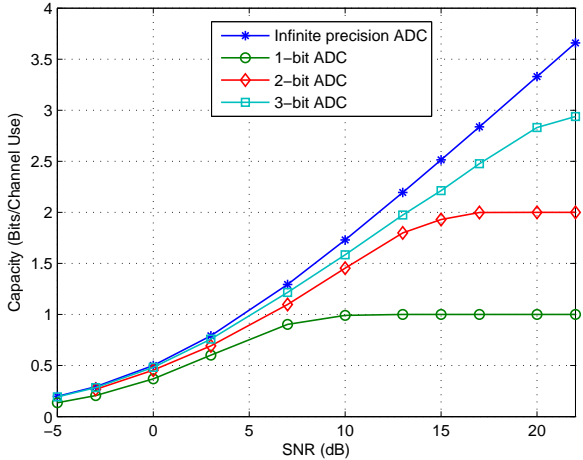}}}
 \end{picture}
  \caption{Capacity with 1-bit, 2-bit, 3-bit, and infinite-precision ADC.}\label{fig:Capacity}
\end{figure}

We now compare the capacity results obtained above with the case
when the receiver ADC has infinite precision. Table
\ref{Tab_Comparison} provides these results, and the
corresponding plots are shown in Fig. \ref{fig:Capacity}. We
observe that at low $\snr$s, low-precision quantization is a
very feasible option. For instance, at -$5$ dB $\snr$, even
$1$-bit receiver quantization achieves $68 \%$ of the capacity
achievable with infinite-precision. $2$-bit quantization at the
same $\snr$ provides as much as $90 \%$ of the
infinite-precision capacity. Such high figures are
understandable, since if noise dominates the message signal,
increasing the quantizer precision beyond a point does not help
much in distinguishing between different signal levels. However,
we surprisingly find that even if we consider moderate to high
$\snr$s, the loss due to low-precision sampling is still very
acceptable. At $10$ dB $\snr$, for example, the corresponding
ratio for $2$-bit quantization is still a very high $85 \%$,
while at $20$ dB, $3$-bit quantization is enough to achieve $85
\%$ of the infinite-precision capacity. Similar encouraging
results have been reported earlier in \cite{Alajaji1, Nossek}
also. However, the input alphabet in these works was taken as
binary to begin with, in which case the good performance with
low-precision output quantization is perhaps less surprising.

On the other hand, if we fix the spectral efficiency to that
attained by an unquantized system at $10$ dB (which is $1.73$
bits/channel use), we find that $2$-bit quantization incurs a
loss of $2.30$ dB (see Table \ref{Tab_Comparison_SE}). From a
practical viewpoint, this penalty in power is more significant
compared to the $15 \%$ loss in spectral efficiency on using
$2$-bit quantization at $10$ dB $\snr$. This suggests, for
example, that the impact of low-precision ADC should be
weathered by a moderate reduction in the spectral efficiency,
rather than by increasing the transmit power. \looseness-1

\begin{table}
\centering
\begin{tabular}{|c|c|c|c|c|c|c|c|c|}
\hline
$\snr$(dB) & $-5$ & $0$ & $5$ & $10$ & $15$ \\
\hline 1-bit ADC & $0.1353$ & $0.3689$ & $0.7684$ & $0.9908$ &  $0.9999$ \\
\hline 2-bit ADC & $0.1792$ & $0.4552$ & $0.8889$ & $1.4731$ &  $1.9304$ \\
\hline 3-bit ADC & $0.1926$ & $0.4817$ & $0.9753$ &  $1.5844$ &  $2.2538$ \\
\hline Unquantized & $0.1982$ & $0.5000$ & $1.0286$ &  $1.7297$ &  $2.5138$ \\
\hline
\end{tabular}
\caption{Capacity (in bits/channel use) at various
$\snr$s.}\label{Tab_Comparison}
\end{table}
\begin{table}
\centering
\begin{tabular}{|c|c|c|c|c|c|c|c|c|}
\hline & \multicolumn{5}{|c|}{Spectral Efficiency (bits per channel use)} \\
\hline & $0.25$ & $0.5$ & $1.0$ & $1.73$ & $2.5$ \\
\hline
\hline 1-bit ADC & $-2.04$ & $1.79$ & $-$ & $-$ &  $-$\\
\hline 2-bit ADC & $-3.32$ & $0.59$ & $6.13$ & $12.30$ & $-$\\
\hline 3-bit ADC & $-3.67$ & $0.23$ & $5.19$ & $11.04$ &  $16.90$\\
\hline Unquantized & $-3.83$ & $0.00$ & $4.77$ & $10.00$ &  $14.91$ \\
\hline
\end{tabular}
\caption{$\snr$ (in dB) required for a given spectral
efficiency.}\label{Tab_Comparison_SE}
\end{table}

\section{Conclusions}\label{sec:Conclusions}
Our Shannon-theoretic investigation indicates the feasibility of
low-precision ADC for designing future high-bandwidth
communication systems such as those operating in UWB and mm-wave
band. The small reduction in spectral efficiency due to
low-precision ADC is acceptable in such systems, given that the
available bandwidth is plentiful. Current research is therefore
focussed on developing ADC-constrained algorithms to perform
receiver tasks such as carrier and timing synchronization,
channel estimation and equalization.

An unresolved technical issue concerns the number of mass points
required to achieve capacity. While we have shown that the
capacity for the AWGN channel with $K$-bin output quantization
is achievable by a discrete input distribution with at most
$K+1$ points, numerical computation of optimal inputs reveals
that $K$ mass points are sufficient. Can this be proven
analytically, at least for symmetric quantizers? Are symmetric
quantizers optimal? 
Another problem for future investigation is whether our result
regarding the optimality of a discrete input can be generalized
to other channel models. Under what conditions is the capacity
of an average power constrained channel with output cardinality
$K$ achievable by a discrete input with at most $K+1$ points?

\bibliographystyle{IEEEbib}
\bibliography{strings,refs,manuals}

\appendices


\end{document}